\documentclass[aps,pra]{revtex4-1}

\usepackage{amsmath}
\usepackage{amssymb}

\begin{document}


\title{A New Loophole in Recent Bell Test Experiments}


\author{Peter Bierhorst}
\email[pbierhor@tulane.edu]
\homepage
\affiliation{Tulane University}

\date{November 17, 2013}

\begin{abstract}
Recent experiments have reached detection efficiencies sufficient to close the detection loophole with photons. Both experiments ran multiple successive trials in fixed measurement configurations, rather than randomly re-setting the measurement configurations before each measurement trial. This opens a new potential loophole for a local hidden variable theory. The loophole invalidates one proposed method of statistical analysis of the experimental results, as demonstrated in this note. Therefore a different analysis will be necessary to definitively assert that these experiments are subject only to the locality loophole.\footnote{Comment, May 2015:  The last sentence of the abstract and the last two paragraphs of this work call for a new form of statistical analysis. A recent work, ``Bell inequalities for continuously emitting sources'' (E. Knill et al., Phys. Rev. A (2015) 91, 032105), has provided such an analysis, and this analysis has been applied to experimental data sets in arXiv:1503.07573 (B.G. Christensen et al.).}
\end{abstract}


\maketitle


\section{Introduction}\label{s:introduction} 

Two recent experiments \cite{guistina:2013,christensen:2013} claim to have closed the detection loophole for photon pairs. As previous experiments have closed the locality loophole for photons (for example, \cite{weihs:1998}), this has generated some reasonable optimism that there may soon be a fully loophole-free test of a Bell inequality. 

The recent experiments are subject to the locality loophole, in the sense that their measurement events were not spacelike separated. This implies that a local hidden variable theory theoretically could have produced the data by exploiting some sort of sub-luminal signalling between the two measurement devices. A natural question then arises: is this the \emph{only} loophole explanation for the data? For instance, the authors of \cite{christensen:2013} point out that the earlier experiment \cite{guistina:2013} may have suffered from some additional loopholes such as the coincidence-time loophole \cite{larsson:2004}. In a reply \cite{larsson:2013}, the first group claims that this can be rectified with a slight modification of the data analysis, but the discussion continues.

To address such matters, one must be clear on the meaning of the following claim:
\begin{equation}\label{e:sentence}
\text{``The experiment is subject only to the locality loophole."}
\end{equation}
We take the meaning of the above statement to be as follows: If the experiment had been performed exactly as described, except for the single change of separating the measuring devices so that all measurement events at one device are spacelike separated from all measurement events at the other device, then the experimental data could not have been produced by any local hidden variable theory except with a vanishingly small probability. This note analyzes the current experimental data according to this notion, concluding that neither team has provided a statistical analysis sufficient to claim (\ref{e:sentence}).

The issue is that both experiments measured multiple successive trials in the same measurement configuration. For instance, \cite{christensen:2013} operated the experiment in a fixed measurement configuration for about 25,000 experimental trials (in 1 second), then randomly re-assigned the measurement settings and measured for another 25,000 trials, repeating this process 4450 times. This is the ``block-measurement" scenario that must be analyzed, and it poses challenges for statistical analysis.

In the online supporting material in \cite{christensen:2013}, it is correctly pointed out that the statistical analyses in the journal versions of \cite{guistina:2013,christensen:2013} are not suitable for bounding the probability of a local hidden variable theory yielding the data \footnote{The journal versions of \cite{guistina:2013,christensen:2013} segment the data and calculate the standard error of the Bell violations over the segments, showing that the results violate the inequality by a large number of standard errors. In addition to ignoring the block measurement design, these error estimates tacitly assume Gaussianity of error terms, an assumption that cannot be supported in a theoretical framework allowing any conceivable local hidden variable theory.}. As an alternative, the authors of \cite{christensen:2013} propose aggregating the blocks of data into groups of size four -- one measurement block for each of the settings configurations -- and analyzing the statitics of these subgroups, or ``cycles." It is asserted that the probability of a local hidden variable theory generating a Bell violation in such a cycle is at most 50\%. Then if the observed number of cycles with Bell violations is much greater than 50\%, a p-value can be assigned to assess the significance of this result.

This seems like a good tactic for directly adressing the block-measurement issue. Unfortunately, the probability of a Bell violation in one of these cycles is not bounded by 50\%, as we will demonstrate in Section \ref{s:blockmeasurement}. Indeed, one can construct a local theory that produces Bell violations at a higher rate than the rate reported in \cite{christensen:2013}. Hence this type of analysis cannot be used to distinguish the experimental results from local theories. It remains an open question if there is a different statistical test that could be applied to the current experimental data to demonstrate the validity of the claim (\ref{e:sentence}).

\section{The Block-Measurement Loophole}\label{s:blockmeasurement} 

In this section, we analyze local hidden variable theories that do not employ any sort of signalling between detectors, so these theories would still apply even if there were complete spacelike separation between the detectors. As previously noted, the measurement settings in \cite{guistina:2013,christensen:2013} are fixed for many successive trials. This provides a local hidden variable with a new dimension to exploit, that would be unavailable if the settings were re-set after each trial. The following table provides an example of a state that a local hidden variable could generate in a block-measurent setting.

\begin{table}[h]\caption{A possible local strategy for a measurement block of 12 trials.}\label{t:astrategy}
\begin{tabular}{ c|c|c|c|c|c|c|c|c|c|c|c|c| }
 \multicolumn{1}{r}{}
  &  \multicolumn{1}{c}{1}
 &  \multicolumn{1}{c}{2}
 &  \multicolumn{1}{c}{3}
  & \multicolumn{1}{c}{4}
    &  \multicolumn{1}{c}{5}
 &  \multicolumn{1}{c}{6}
 &  \multicolumn{1}{c}{7}
  & \multicolumn{1}{c}{8} 
 &  \multicolumn{1}{c}{9}
 &  \multicolumn{1}{c}{10} 
   & \multicolumn{1}{c}{11} 
 &  \multicolumn{1}{c}{12}  \\
 \cline{2-13}
 $a$ & + & + & + & + & + & + & + & + & + & 0 & 0 & 0  \\
 \cline{2-13}
 $a'$ & + & + & + & 0 & 0 & + & + & + & + & 0 & 0 & 0 \\
  \cline{2-13}
 $b$ & + & + & + & + & + & + & + & + & + & 0 & 0 & 0 \\
 \cline{2-13}
 $b'$ & 0 & 0 & 0 & + & + & + & + & + & + & 0 & 0 & 0 \\
 \cline{2-13}
 \end{tabular}
\end{table}

It is worthwhile to study this example and how it relates to the Clauser-Horne inequality. In the form relevant to our discussion, the Clauser-Horne inequality states that 
\begin{equation}\label{e:CHforBML}
P(++|ab)-P(+0|ab')-P(0+|a'b)-P(++|a'b')\le 0.
\end{equation}
We can see that if the state in Table \ref{t:astrategy} is measured in the $ab$ basis, it will produce 9 ``++" counts and 3 ``00" counts. Thus the estimate of $P(++|ab)$ would be ${9\over 12}$. Now, this local strategy - like all local strategies - must invariably obey the inequality (\ref{e:CHforBML}). Thus for each ``++" count in the $ab$ basis, there is a contribution to one of the negative terms in (\ref{e:CHforBML}). In Table \ref{t:astrategy}, the first three ++$ab$ counts are balanced by three +0 counts that would appear if the state were measured in the $ab'$ basis. The following two ++$ab$ counts are balanced by two 0+ counts in the $a'b$ basis, and the last four are balanced by four ++ counts in the $a'b'$ basis. Hence this state yields  
\begin{equation}\label{e:examplestate}
P(++|ab)= {9\over 12} \quad P(+0|ab')= {3\over 12} \quad P(0+|a'b)= {2\over 12} \quad P(++|a'b')= {4\over 12}
\end{equation}
and the Clauser-Horne inequality is satisfied. Now, in an experiment like \cite{christensen:2013} there are 25,000 trials instead of 12. In such a setting, a local hidden variable can mimic any potential value for $P(++|ab)$ to many decimal points of precision and can divide up that probability between the three negative terms in (\ref{e:CHforBML}) in any varying amount that saturates the inequality.

Now, suppose we were to measure many such blocks, each with randomized measurement settings. Then suppose we collect these blocks into groups of four, one with each measurement setting, as described in \cite{christensen:2013}. It is asserted in that, under a local theory, the probability of a cycle violating the bell inequality (\ref{e:CHforBML}) is at most 50\%.

This does not appear to be the case. Here is an example of a local strategy, exploiting the block measurement loophole, that produces Bell violations in more than 50\% of cycles. Consider the following three distributions:
\begin{equation}\label{e:counter1a}
P(++|ab)= {24\over 72} \quad P(+0|ab')= {8\over 72} \quad P(0+|a'b)= {8\over 72} \quad P(++|a'b')= {8\over 72}
\end{equation}
\begin{equation}\label{e:counter1b}
P(++|ab)= {18\over 72} \quad P(+0|ab')= {6\over 72} \quad P(0+|a'b)= {6\over 72} \quad P(++|a'b')= {6\over 72}
\end{equation}
\begin{equation}\label{e:counter1c}
P(++|ab)= {3\over 72} \quad P(+0|ab')= {1\over 72} \quad P(0+|a'b)= {1\over 72} \quad P(++|a'b')= {1\over 72}
\end{equation}
Following the example in Table \ref{t:astrategy}, one can construct local states that can produce each of these probability distributions. Now, suppose the local hidden variable theory sends states yielding distributions (\ref{e:counter1a}), (\ref{e:counter1b}), and (\ref{e:counter1c}) with respective probabilities ${1\over 2}$, ${1\over 4}$, and ${1\over 4}$. Then we can calculate the probability of a Bell violation, as follows. In a given cycle, there is a ${1\over 2}$ probability that the state was (\ref{e:counter1a}) when the $ab$ setting was measured. If this is the case, then there will be a violation of (\ref{e:CHforBML}) if at least one of the other settings was measured in state (\ref{e:counter1b}) or (\ref{e:counter1c}). This yields
$$
P\big(ab = \text{(\ref{e:counter1a})} \big)\cdot P\big(\text{Bell Violation} \big| ab= (\ref{e:counter1a}) \big) = {1\over 2}\cdot \bigg[ 1- {1\overwithdelims() 2}^3\bigg] = {7\over 16}.
$$
Moreover, this is not the only way that a Bell violation can be obtained. If the $ab$ setting was measured when the state was \text{(\ref{e:counter1b})}, which occurs with probability ${1\over 4}$, a Bell violation will be seen if at least one of the other settings was measured in state (\ref{e:counter1c}). This yields 
$$
P\big(ab= \text{(\ref{e:counter1b})} \big)\cdot P\big(\text{Bell Violation} \big| ab= (\ref{e:counter1b}) \big) = {1\over 4}\cdot \bigg[ 1- {3\overwithdelims() 4}^3\bigg] = {37\over 256}.
$$
The total probability of a Bell violation is the sum of these two probabilities, which is ${149\over 256}\simeq58.2\%$. The analysis of \cite{christensen:2013} saw 394 violations in 650 cycles (60.6\% Bell violations), a figure which is a significant deviation from 50\%, but not from 58.2\%. Indeed, it turns out that a local hidden variable can actually do better than 60.6\%. Let $\rho$ be any number between 0 and 1, and choose any positive $\epsilon$ smaller than ${\rho\over 40}$. Then consider the following four local strategies:
\begin{equation}\label{e:counter2a}
P(++|ab)= \rho \quad P(+0|ab')= {\rho\over 3} \quad P(0+|a'b)= {\rho \over 3} \quad P(++|a'b')= {\rho\over 3}
\end{equation}
\begin{equation}\label{e:counter2b}
P(++|ab)= \rho - 3\epsilon \quad P(+0|ab')= {\rho\over 3}-\epsilon \quad P(0+|a'b)= {\rho \over 3}-\epsilon \quad P(++|a'b')= {\rho\over 3}-\epsilon
\end{equation}
\begin{equation}\label{e:counter2c}
P(++|ab)= \rho - 12\epsilon \quad P(+0|ab')= {\rho\over 3}-4\epsilon \quad P(0+|a'b)= {\rho \over 3}-4\epsilon \quad P(++|a'b')= {\rho\over 3}-4\epsilon
\end{equation}
\begin{equation}\label{e:counter2d}
P(++|ab)= \rho - 39\epsilon \quad P(+0|ab')= {\rho\over 3}-13\epsilon \quad P(0+|a'b)= {\rho \over 3}-13\epsilon \quad P(++|a'b')= {\rho\over 3}-13\epsilon.
\end{equation}
A local hidden variable theory sending these states with respective probabilities ${2\over 5}$, ${1\over 4}$, ${1\over 5}$, and ${3\over 20}$ will violate the Clauser-Horne inequality with probability exceeding 63\%. 

How is this possible? The strategies described above violate the Clauser-Horne inequality with large probability, but since they are local, they still must obey it in expectation. This means frequent small violations of the inequality (\ref{e:CHforBML}) are balanced out by less frequent, but larger, deviations in the other direction (results satisfying the inequality by a large margin). Nevertheless, these examples eliminate the strategy of using ``probability of a Bell violation in a given cycle" as a test parameter for distinguishing quantum theories from local theories with the data of \cite{christensen:2013}.

\section{Types of Spacelike Separation}\label{s:types}

The local hidden variable models described in the previous section are applicable even if the strongest notion of locality is enforced. We call this \emph{Type 1} separation: the last measurement event at one end of the experiment takes place before a not-faster-than-light signal could have arrived from the first measurement event at the other end of the experiment. 

A natural question arises: how far apart would the detectors have to be to achieve this type of separation? The duration of the entire experiments in \cite{guistina:2013,christensen:2013} were about 20 minutes and 75 minutes, respectively. The diameter of Earth's orbit around the Sun is roughly 16 light-minutes, so this is the sort of distance that would be necessary to enforce Type 1 separation.

A second, weaker type of separation would be as follows: a block of measurements in a particular setting may be able to receive sub-luminal signals containing information about previous measurement blocks from the other end of the experiment, but within a measurement block, even the last trial in the block is spacelike separated from every trial in the same block at the other side of the experiment. We call this \emph{Type 2} separation. The length of the measurement blocks in the two aforementioned experiments are 1 minute and 1 second, respectively; light takes about 1.3 seconds to travel from the Earth to the Moon.

If one exploits the block measurement loophole in a Type 2 separation scenario, the results can be even more pathological, if we also allow the hidden variable access to sub-luminal signaling. For a simple illustration, suppose the local theory starts by sending state (\ref{e:counter1b}) for the first measurement block. Since only Type 2 separation is enforced, we can suppose the local theory has access to complete information to the settings and outcomes of the first $n-1$ blocks prior to sending a state for the $n$th block. Thus the theory could keep sending state (\ref{e:counter1b}) until all but one of the four setting configurations has been measured. At this point, if the missing configuration is $ab$, the local theory now starts sending state (\ref{e:counter1a}), and if the missing configuration is $ab'$, $a'b$, or $a'b'$, the theory starts sending state (\ref{e:counter1c}). This will produce a Bell violation on the first measurement cycle with probability 1.

A third, weakest type of separation -- \emph{Type 3}  separation -- would be as follows: separate the experiments just enough so that simultaneous trials are space-like separated, but trials occurring later in a block can possibly receive sub-luminal signals from trials occuring earlier within the same block on the other side of the experiment. The experiment \cite{christensen:2013} was the only one of the two experiments to have intrinsically well-defined trials within blocks, which lasted .04 ms. This is the time it takes light to travel 12 km.

Type 3 separation is the easiest type of separation to achieve. It is also fairly clear that if only Type 3 separation is enforced, a local hidden variable could generate just about any desired distribution by employing the following strategy: at the beginning of a block of measurements, generate a few throwaway outcomes at detector A, waiting for information about the setting at detector B to arrive. Once that information arrives, the hidden variable has full information about all the settings, and can simulate any desired distribution for the rest of the block - which is 25,000 trials long, in the case of \cite{christensen:2013}.

Therefore, if separations on the order of a few kilometers are all that can be achieved, the experimental design would have to be changed. One way to eliminate the block measurement issue would be to employ superfast random setting-switchers so that each trial has randomized settings uncorrelated to the settings of other trials. Another related method would be to analyze only the first trial after each switching of the settings, throwing away the results of all other trials. As shown elsewhere \cite{bierhorst:inprep}, the data from the experiments  \cite{guistina:2013,christensen:2013} is indeed sufficient to rule out all loopholes (minus the locality loophole), if the detector settings are assumed to have been randomly toggled prior to each measurement event. Insofar as it is reasonable to believe the data would be the same if the settings were indeed randomized for each measurement event, the prognosis for success for such an experiment would be favorable.


The claim (\ref{e:sentence}) may yet be salvaged. It has not been demonstrated that it is possible to replicate all of the data in the experiments \cite{guistina:2013,christensen:2013} by exploiting local strategies of the form described in Section \ref{s:blockmeasurement}. Rather, it has only been shown that the current statistical analyses are not capable of distinguishing the measurement data from local theories. There may be a yet-undescribed statistical test that distinguishes the experimental data from any local hidden variable theory that does not exploit signaling between the detectors. If this could be done, then the claim (\ref{e:sentence}) could be made.

This is a worthy goal, as the theoretical implications are tangible. If it can be shown that the experiments \cite{guistina:2013,christensen:2013} are indeed subject to \emph{only} the locality loophole, then the local theorist is confronted with a truly daunting task if he must explain the behavior of photons. For small distances, one could say any local theory must exploit signalling between the two members of the photon pair to display quantum correlations. As the photons start to get farther apart, the theory must toggle to a different mechanism to mimic quantum correlations, now exploiting non-detection to evade certain setting configurations. It would be a notable achievement to restrict the class of local theories for photons in this manner.

\bibliographystyle{unsrt}
\bibliography{bibliography}

\begin{thebibliography}{1}

\bibitem{guistina:2013}
M.~Giustina et~al.
\newblock Bell violation using entangled photons without the fair-sampling
  assumption.
\newblock {\em Nature}, 497:227--30, 2013.

\bibitem{christensen:2013}
B.~G. Christensen et~al.
\newblock Detection-loophole-free test of quantum nonlocality, and
  applications.
\newblock {\em Phys. Rev. Lett.}, 111:130406, Sep 2013.

\bibitem{weihs:1998}
G.~Weihs, T.~Jennewein, C.~Simon, H.~Weinfurter, and A.~Zeilinger.
\newblock Violation of {B}ell's inequality under strict {E}instein locality
  conditions.
\newblock {\em Phys. Rev. Lett.}, 81:5039--43, Dec 1998.

\bibitem{larsson:2004}
J.-\AA Larsson and R.~D. Gill.
\newblock Bell's inequality and the coincidence-time loophole.
\newblock {\em Europhys. Lett.}, 67(5):707--13, 2004.

\bibitem{larsson:2013}
J.-\AA Larsson, M.~Giustina, J.~Kofler, B.~Wittmann, R.~Ursin, and S.~Ramelow.
\newblock Bell violation with entangled photons, free of the coincidence time
  loophole.
\newblock 2013.
\newblock arXiv:1309.0712 [quant-ph].

\bibitem{Note1}
The journal versions of \cite {guistina:2013,christensen:2013} segment the data
  and calculate the standard error of the Bell violations over the segments,
  showing that the results violate the inequality by a large number of standard
  errors. In addition to ignoring the block measurement design, these error
  estimates tacitly assume Gaussianity of error terms, an assumption that
  cannot be supported in a theoretical framework allowing any conceivable local
  hidden variable theory.

\bibitem{bierhorst:inprep}
P.~Bierhorst.
\newblock in preparation.

\end{thebibliography}

\end{document}